\documentclass[aps,prc,twocolumn,amsmath,amssymb,superscriptaddress]{revtex4-1}
\usepackage{graphicx}
\usepackage{hyperref}
\usepackage{lineno}
\usepackage{color}
\linenumbers
\newcommand{\be}{\begin{equation}}
\newcommand{\ee}{\end{equation}}
\newcommand{\beq}{\begin{eqnarray}}
\newcommand{\eeq}{\end{eqnarray}}
\newcommand{\ba}{\begin{array}}
\newcommand{\ea}{\end{array}}

\begin{document}

\title{Threshold Upsilon-meson Photoproduction at EIC and EicC}

\date{\today}

\author{\mbox{Igor~I.~Strakovsky}}
\altaffiliation{Corresponding author: \texttt{igor@gwu.edu}}
\affiliation{Institute for Nuclear Studies, Department of Physics, The 
	George Washington University, Washington, DC 20052, USA}

\author{\mbox{William~J.~Briscoe}}
\affiliation{Institute for Nuclear Studies, Department of Physics, The
        George Washington University, Washington, DC 20052, USA}

\author{\mbox{Lubomir~Pentchev}}
\affiliation{Thomas Jefferson National Accelerator Facility, Newport 
	News, Virginia 23606, USA}

\author{\mbox{Axel~Schmidt}}
\affiliation{Institute for Nuclear Studies, Department of Physics, The
        George Washington University, Washington, DC 20052, USA}

\noaffiliation

\begin{abstract}
High-accuracy $\Upsilon$-meson photoproduction data from EIC and EicC experiments will allow the measurement of the near-threshold total cross section of the reaction $\gamma p\to\Upsilon p$, from which the absolute value of the $\Upsilon p$ scattering length, $|\alpha_{\Upsilon p}|$, can be extracted using a Vector-Meson Dominance model. For this evaluation, we used $\Upsilon$-meson photoproduction quasi-data from the QCD approach (the production amplitude can be factorized in terms of gluonic generalized parton distributions and the quarkonium distribution amplitude). A comparative analysis of $|\alpha_{\Upsilon p}|$ with the recently determined scattering lengths for $\omega p$, $\phi p$, and $J/\psi p$ using the A2, CLAS, and GlueX experimental data are performed. The role of the ``young'' vector-meson effect is evaluated.
\end{abstract}

\maketitle

\section{Introduction}
\label{sec:intro} 

In 1977, the observation of an enhancement at 9.5~GeV in the di-muon mass spectrum produced in 400-GeV proton-nucleus collisions at Fermi Lab resulted in a new vector-meson ($V$)~\cite{Herb:1977ek}, then called $\Upsilon(1S)(9460)$.

Compared to other mesons, vector-mesons can be measured to very high precision.  This stems from the fact that vector-mesons have the same quantum numbers as the photon  $I^G(J^{PC}) = 0^-(1^{-~-})$.

The photoproduction of $J/\psi(1S)$ and $\Upsilon(1S)$ are among the key reactions that will be measured at the electron-ion colliders EIC hosted by the Brookhaven National Laboratory (BNL)~\cite{AbdulKhalek:2021gbh} and EicC at the High Intensity heavy-ion Accelerator Facility (HIAF) in China~\cite{Anderle:2021wcy}. Charmonium and bottomonium are important not only for understanding the interaction mechanisms of the photoproduction of the heavy vector-mesons, but also for probing the gluonic properties of the nucleon. The large statistics of the exclusive $J/\psi(1S)$ and $\Upsilon(1S)$ production data at the hard scale are very helpful in extracting the generalized parton distribution (GPD) of the gluon. These measurements will advance our understanding of QCD which governs the properties of hadrons and the interactions involving hadrons. 

 Exclusive vector meson photoproduction can also shed light on the bound meson-nucleon system, since the generated charmonium and bottomonium interact with an intact nucleon. We point out that the bottomonium, $\Upsilon(1S)$, can be measured using real photons at EIC and EicC, where the quality of the expected data near-threshold will give access to a variety of interesting physics aspects, \textit{e.g.}, trace anomaly, pentaquarks, cusp effects, vector-meson-nucleon scattering length, and so on. The main object of this note is to estimate the magnitude of the absolute value of the $\Upsilon p$ scattering length, $|\alpha_{\Upsilon p}|$, using quasi data generated from the QCD model of Ref.~\cite{Guo:2021ibg}, and compare it with the results for the other vector mesons. Our analysis is based also on the Vector-Meson Dominance (VMD) model~\cite{GellMann:1961tg,Sakurai:1969} relying on the transparent current-field identities of Kroll, Lee, and Zumino~\cite{Kroll:1967it}. The VMD model can be used for a variety of qualitative estimates of observables in vector-meson photoproduction~\cite{Lutz:2001mi,Titov:2007xb} at least as the first step towards their more extended theoretical studies. The use of the VMD model in case of the $J/\psi$ and $\Upsilon$ requires special attention due to the heavy mass of these vector-mesons. For the critical review of the VMD model, we refer to papers of Boreskov and Ioffe~\cite{Boreskov:1976dj} and Kopeliovich and co-workers~\cite{Kopeliovich:2017jpy} and references therein. Recently, in Ref.~\cite{Xu:2021mju}, the VMD approach has been examined based on Dyson-Schwinger equations (DSE).

On the basis of recent threshold measurements of the photoproduction of three vector mesons off the proton by the A2 (MAMI), CLAS (JLab), and GlueX (JLab) Collaborations, one can determine vector-meson--proton scattering lengths (SLs) using the VMD model~\cite{Strakovsky:2014wja,Strakovsky:2020uqs,Strakovsky:2019bev}. This results in
\begin{equation}
	    |\alpha_{J/\psi p}|~\ll~|\alpha_{\phi p}|~\ll~|\alpha_{\omega p}|, 
        \label{eq:eqa}
\end{equation}
which indicates that the proton is more transparent for the $\phi$-meson compared to the $\omega$-meson, and at the same time it is much less transparent for the $\phi$-meson than for the $J/\psi$-meson. Due to the small size of ``young'' vector mesons relative to those that have had time to fully form, scattering lengths determined phenomenologically in the near threshold photoproduction are smaller. Recall that when the photon produces a vector meson, $V$, it first creates a $q\bar q$ pair in point-like configuration. Near the threshold, this pair lacks sufficient time to form the complete wave function of the vector meson; that is, the proton interacts with the ``young'' (undressed) vector meson whose size is smaller than that of the ``old'' one participating in the elastic $Vp\to Vp$ scattering. Therefore, one observes stronger suppression for the vector-meson--proton interaction~\cite{Feinberg:1980yu}.

\section{Exclusive Vector-Meson Photoproduction at the EIC}
\label{sec:EIC}

Exclusive vector-meson photoproduction is one of the key physics measurements for the EIC as discussed in the Yellow Report~\cite{AbdulKhalek:2021gbh}. The proposed design for EIC is to collide electrons with energy of $E_e = 5$--$18$~GeV and protons with energy of $T_p = 41$--$275$~GeV. The proposed EIC detector has a number of specific features that will enable photo-production measurements, and to ensure exclusivity. First, the energy of quasi-real photons will be determined by tagger detectors in the so-called ``far-backward'' region, i.e., in the direction of the electron beam. Tagger detectors placed approximately 24~m and 37~m from the interaction point can cover a low-$Q^2$ acceptance better than $10^{-7}$~GeV$^2/c^2$. Produced $\Upsilon(1S)$ mesons will be reconstructed from their leptonic decays, e.g., to an $e^+e^-$ pair. The proposed EIC detector aims for momentum resolution sufficient to cleanly separate the $\Upsilon$-states, a $\Delta p/p$ of better than 1\% in the 4--10 GeV/$c$ momentum range. Finally, exclusivity can be ensured by the detection of the final state proton, which, at threshold, will largely travel to the so-called ``far-forward'' region of the detector. The far-forward region, covering angles within approximately 13~mrad from the proton beam line, will be instrumented with a series of tracking detectors including Roman Pots, to track charged particles with slightly different magnetic rigidities than beam protons. Thus threshold $\Upsilon$ production will require the combination of the far-backward, far-forward and main EIC detectors in order to reconstruct the full event and ensure exlusivity.

\section{Scattering Length for Upsilon - Proton}
\label{sec:SL}

\begin{figure}[htb]
\begin{center}
\includegraphics[height=1.2in, keepaspectratio]{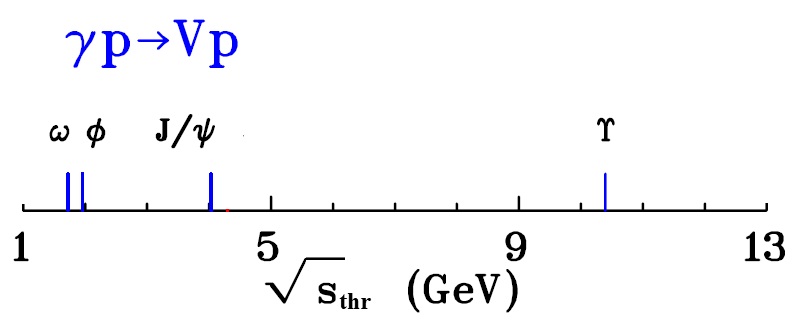}
\end{center}

\vspace{-7mm}
\caption{Thresholds of meson photoproduction off the proton. Blue vertical lines are for                    vector-mesons including charmonium and b-quarkonium.}  
	    \label{fig:figx}
\end{figure}

The total cross section of a binary reaction $ab\to cd$ with particle masses $m_a + M_b < m_c + M_d$ can be written as 
\begin{equation}
	    \sigma_t = \frac{q}{k}\cdot F(q,s),
	    \label{eq:eq1} 
\end{equation}
where $s$ is the square of the total center-of-mass energy, $k$ is center-of-mass momentum of $a$ (and $b$), and $q$ is the center-of-mass momentum of $c$ (and $d$). In photoproduction, i.e., when $m_a = 0$, $k = (s - M_b^2)/(2\cdot\sqrt{s})$. Vector-meson kinematical parameters for the vector-meson photoproduction off the proton, i.e., $\gamma p\to Vp$, are given in Table~\ref{tbl:tab1} and Figure~\ref{fig:figx}, where $m_V$ is the vector-meson mass, $s_{thr}$ is the value of $s$ at threshold, $k_{thr}$ is the value of $k$ at threshold, and $E_{thr}$ is the photon energy at threshold in the frame where the proton is initially at rest.

The factor $F(q,s)$ in Eq.~\ref{eq:eq1} is proportional to the square of the invariant amplitude and does not vanish at threshold, i.e., when $q\to 0$ and $k\to k_{thr}$, but instead approaches a constant value. Thus, near threshold, $\sigma_t\to 0$ and is at least proportional to $q$.
\begin{table}[htb!]

\centering
\protect\caption{Kinematical parameters for the vector-meson photoproduction off the proton at 
        thresholds~\protect\cite{Zyla:2020zbs}.}
\vspace{2mm}
{%
\begin{tabular}{|c|c|c|c|c|}
\hline
Vector-        & $m_V$    &$\sqrt{s_{thr}}$& $E_{thr}$& $k_{thr}$\\
Meson          & (MeV)    &   (MeV)        &  (MeV)   & (MeV/$c$)\\
\hline
$\omega(782)$  &  782.65  &  1720.9        &  1109.1  &  604.7   \\
$\phi(1020)$   & 1019.461 &  1957.7        &  1573.3  &  754.0   \\
$J/\psi(1S)$   & 3096.900 &  4035.2        &  8207.8  & 1908.5   \\
$\Upsilon(1S)$ & 9460.30  & 10398.6        & 57152.9  & 5156.9   \\
\hline
\end{tabular}} \label{tbl:tab1}
\end{table}

Traditionally, the $\sigma_t$ behavior of a near-threshold binary inelastic reaction is described as a series of odd powers in $q$ (see, for instance, Ref.~\cite{Strakovsky:2014wja}). 
In the energy range under study, we use:
\begin{equation}
        \sigma_t(q) = b_1\cdot q + b_3\cdot q^3 + b_5\cdot q^5,
        \label{eq:eq2}
\end{equation}
which assumes contributions from only the lowest $S$-, $P$-, and $D$-waves. Very close to threshold, the higher-order terms can be neglected and the linear term is determined by the $S$-wave only with a total spin of $1/2$ and/or $3/2$.  

\begin{figure}[htb]
\begin{center}
\includegraphics[height=1.1in, keepaspectratio]{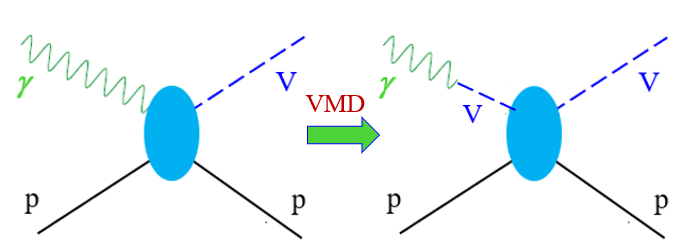}
\end{center}
\vspace{-7mm}

\caption{Schematic diagrams of vector-meson photoproduction (left) and the VMD model (right) in the energy     
        region at threshold experiments.} 
        \label{fig:figy}
\end{figure}

For the evaluation of the absolute value of the vector-meson--proton SL, we apply the commonly used and effective VMD approach (Fig.~\ref{fig:figy}), which links the near-threshold cross sections of the vector-meson photoproduction, $\gamma p\to Vp$, and the elastic scattering, $Vp\to Vp$, processes via~\cite{Titov:2007xb}:
\begin{eqnarray}
        \frac{d\sigma^{\gamma p\to Vp}}{d\Omega}|_{\rm thr}
        & = &\frac{q}{k} \cdot \frac{1}{64\pi}\cdot |T^{\gamma p\to Vp}|^2\nonumber\\
        & = &\frac{q}{k} \cdot \frac{\pi\alpha}{g_{V}^2}\cdot
        \frac{d\sigma^{Vp\to Vp}}{d\Omega}|_{\rm thr}\nonumber\\
        & = &\frac{q}{k} \cdot \frac{\pi\alpha}{g_{V}^2}\cdot
        |\alpha_{Vp}|^2,
        \label{eq:eq3}
\end{eqnarray}
where $T^{\gamma p\to Vp}$ is the invariant amplitude of the vector-meson photoproduction, $\alpha$ is the fine-structure constant, and $g_V$ is the VMD coupling constant, related to the vector-meson electromagnetic (EM) decay width $\Gamma_{V\to e^+e^-}$
\begin{equation}
        g_V^2 = \frac{\pi\cdot\alpha^2\cdot m_V}{3\cdot\Gamma(V\to e^+e^-)},
        \label{eq:eq4}
\end{equation}
where $m_V$ is the vector-meson mass.  

\vspace{1cm}
\vspace{-12mm}

\vspace{5mm}
Combining Eq.~(\ref{eq:eq2}) (which is also valid for $\omega$- $\phi$-, and $J/\psi$-photoproduction~\cite{Strakovsky:2014wja,Strakovsky:2019bev,Strakovsky:2020uqs}) and Eq.~(\ref{eq:eq3}) with Eq.~(\ref{eq:eq4}), one can express the absolute value of the SL as a product of the pure EM, VMD-motivated kinematic factor
\begin{eqnarray}
	    B_V^2 = \frac{{\alpha\cdot m_V\cdot k}}
	    {{12\pi\cdot\Gamma(V\to e^+e^-)}} 
	    \label{eq:eq5}
\end{eqnarray}
and the factor 
\begin{eqnarray}
	    h_{Vp} = \sqrt{b_1} 
        \label{eq:eq6}
\end{eqnarray}
that is determined by an interplay of strong (hadronic) and EM dynamics as
\begin{eqnarray}
        |\alpha_{Vp}| = B_V\cdot h_{Vp}.
        \label{eq:eq7}
\end{eqnarray}
Numerical values of both $g_V$ and $B_V$ are given in Table~\ref{tbl:tab2}. Let us note that these EM quantities $B_V$ for each vector-meson are close to each other except for $\Upsilon$.
\begin{table}[htb!]

\centering 
\protect\caption{Vector-meson EM properties. The decay $\Gamma(V\to e^+e^-)$ from
        PDG2020~\protect\cite{Zyla:2020zbs} (2nd column), $g_V$ was calculated using 
        Eq.~(\ref{eq:eq4}) (3rd column) and $B_V$ using Eq.~(\ref{eq:eq5}) (4th column).} 
\vspace{2mm}
{%
\begin{tabular}{|c|c|c|c|c|}
\hline
Vector-        & $\Gamma(V\to e^+e^-)$ & $g_V$     &       $B_V$ \\
Meson          &        (keV)          &           &   (MeV$^{1/2}$) \\
\hline
$\omega(782)$  & 0.60$\pm$0.02   &  8.53$\pm$0.14  &  390.49$\pm$6.35 \\
$\phi(1020)$   & 1.27$\pm$0.04   &  6.69$\pm$0.10  &  342.50$\pm$5.27 \\
$J/\psi(1S)$   & 5.53$\pm$0.10   &  5.59$\pm$0.05  &  454.92$\pm$4.06 \\
$\Upsilon(1S)$ & 1.340$\pm$0.018 & 19.85$\pm$1.21  & 2654.96$\pm$162.15 \\
\hline
\end{tabular}} \label{tbl:tab2}
\end{table}

\begin{figure}[htb]
\begin{center}
\includegraphics[height=2.5in, keepaspectratio]{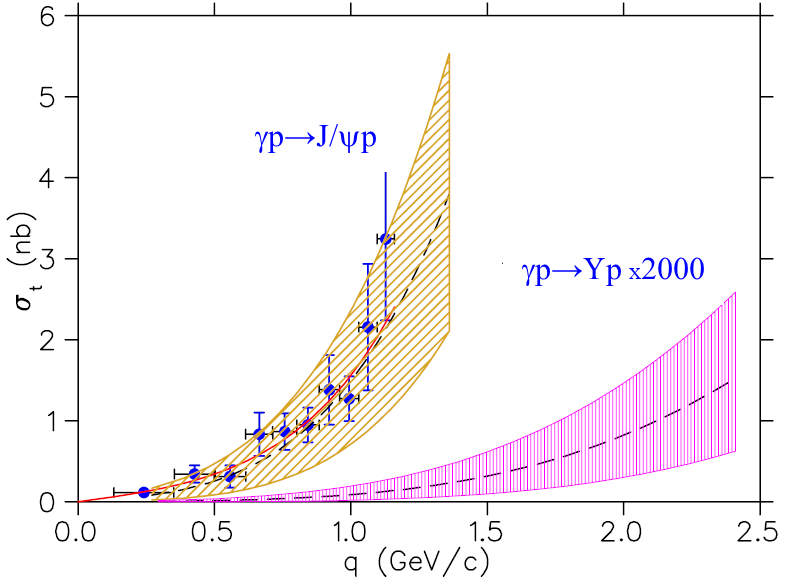}
\end{center}

\vspace{-7mm}
\caption{The total $\gamma p\to J/\psi p$ and $\gamma p\to\Upsilon p$ cross sections $\sigma_t$. 
        GlueX threshold $J/\psi$ photoproduction data~\protect\cite{Ali:2019lzf} shown by blue filled circles and the best-fit results are from Ref.~\protect\cite{Strakovsky:2019bev} and shown by red solid curve. Theoretical fit of the GlueX data from Ref.~\protect\cite{Guo:2021ibg} is shown by yellow shaded band at 95\% C.L. Theoretical predictions for $\Upsilon$ photoproduction~\protect\cite{Guo:2021ibg} are shown by magenta shaded band.}  \label{fig:fig1}
\end{figure}

Figure~\ref{fig:fig1} shows the fit of the QCD model of Ref.~\cite{Guo:2021ibg} to the GlueX $J/\psi$ photoproduction data~\cite{Ali:2019lzf} and the prediction of this model for the $\Upsilon$ photoproduction total cross section. Our phenomenological fit of the GlueX data using Eq.~(\ref{eq:eq2}) is shown, as well. This is a rather general model assuming factorization in terms of gluonic Generalized Parton Distributions (GPD) and quarkonium wave function on one side and hard quark-gluon interaction on the other side.  This work extends the validity of the factorization, as studied previously in Ref.~\cite{Ivanov:2004vd} for high energies, down to the threshold region in leading order and in the case of a heavy quark mass.

\begin{figure}[htb]
\begin{center}
\includegraphics[height=2.5in, keepaspectratio]{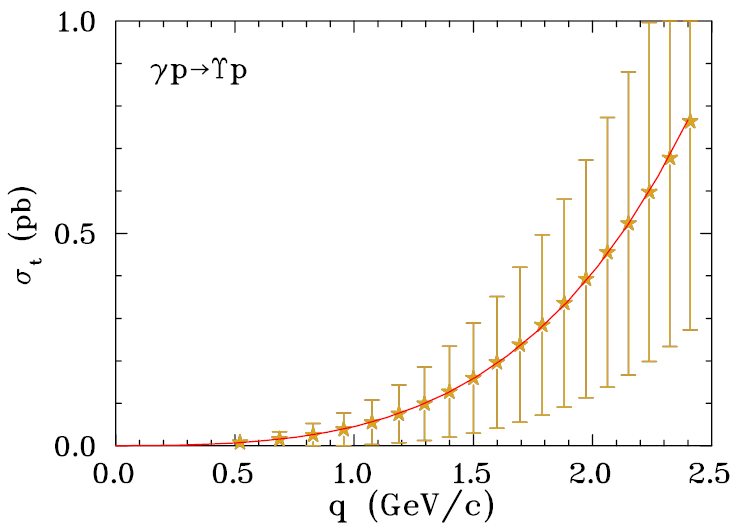}
\end{center}

\vspace{-7mm}
\caption{The total $\gamma p\to J/\psi p$ cross sections $\sigma_t$. Quasi-data from 
        Ref.~\protect\cite{Guo:2021ibg} ($q_{min}$ = 521~MeV/$c$) shown by yellow filled stars 
        The phenomenological best fit for quasi cross section~\protect\cite{Guo:2021ibg} is shown by red solid curve. The uncertainties for quasi-data were taken from Ref.~\protect\cite{Guo:2021ibg}. 
        }  
    \label{fig:fig2}
\end{figure}

\begin{figure}[htb]
\begin{center}
\includegraphics[height=2.5in, keepaspectratio]{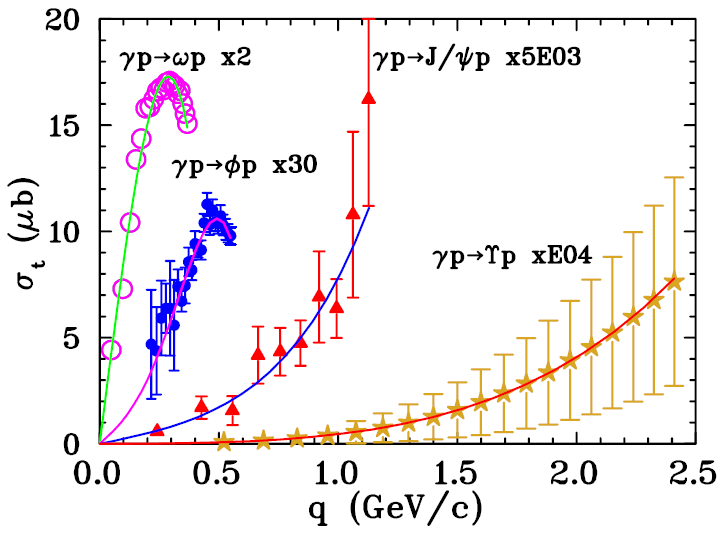}
\end{center}

\vspace{-7mm}
\caption{The total $\gamma p\to Vp$ cross section $\sigma_t$ derived from the A2 (magenta open 
        circles)~\protect\cite{Strakovsky:2014wja}, CLAS (blue filled circles)~\protect\cite{Strakovsky:2020uqs}, and GlueX (red filled triangles)~\protect\cite{Strakovsky:2019bev}, data, and EIC/EicC (yellow filled stars) quasi-data is shown as a function of the center-of-mass momentum $q$ of the final-state particles.  The vertical (horizontal) error bars represent the total uncertainties of the data summing statistical and systematic uncertainties in quadrature (energy binning). Solid curves are the fit of the data with Eq.~(\protect\ref{eq:eq2}).}  
  \label{fig:fig3}
\end{figure}

We have generated quasi-data (Fig.\ref{fig:fig2}) for the $\Upsilon$ cross sections from this model~\cite{Guo:2021ibg} and fit it with Eq.~(\ref{eq:eq2}) in order to extrapolate the data to the threshold. The exact $q_{min}$ attainable in an $ep$ collider experiment will depend heavily on the exact placement of the low-$Q^2$ tagging detectors. For the purposes of this note, we assume that the detector will have comparable coverage to the EIC Yellow Report detector~\cite{AbdulKhalek:2021gbh}, where the backward calorimeter coverage down to a pseudo-rapidity of $-4$  would allow $q_{min}$ for $\Upsilon$ photoproduction to be as low as $\approx 500$~MeV/c. Further optimization of the low-$Q^2$ taggers may allow an even smaller $q_{min}$ to be achieved. Gryniuk and co-workers assumed a total integrated luminosity of 100~$fb^{-1}$ for the $\Upsilon$ photoproduction at EIC, which corresponds to 116~days of the beam with the $10^{34}~cm^{-2}~s^{-1}$, for Monte Carlo calculations~\cite{Gryniuk:2020mlh}. As Guo and co-workers~\cite{Guo:2021ibg} pointed out, the large mass of the $\Upsilon$-meson implies that the calculations in the heavy meson limit works better for the $\Upsilon$ photoproduction, as there are fewer theoretical uncertainties from higher-order corrections. 
\begin{table*}[htb!]

\centering
\protect\caption{The 2nd column showed the minimal momentum $q_{min}$ for vector-mesons in photoproduction 
        experiments and the source of data. The linear term (3rd column) of the best-fit of the total cross section data using Eq.~(\protect\ref{eq:eq2}). The errors represent the total uncertainties (summing statistical and systematic uncertainties in quadrature. The 4th column showed vector-meson--proton SLs.}
\vspace{2mm}
{%
\begin{tabular}{|c|c|c|c|}
\hline
Vector-          &   $q_{min}$                           &      $b_1$        & $|\alpha_{Vp}|$ \\
Meson            &   (MeV/$c$)                           & ($\mu b$/MeV/$c$) &  (fm) \\
\hline
$\omega(782)$    &  49~\protect\cite{Strakovsky:2014wja} & (0.44$\pm$0.01)$\times 
10^{-1}$~\protect\cite{Strakovsky:2014wja} & 0.82$\pm$0.03~\protect\cite{Strakovsky:2014wja} \\
$\phi(1020)$     & 216~\protect\cite{Dey:2014tfa}        & (0.34$\pm$0.12)$\times 
10^{-3}$~\protect\cite{Strakovsky:2020uqs} & 0.063$\pm$0.010~\protect\cite{Strakovsky:2020uqs} \\
$J/\psi(1S)$     & 230~\protect\cite{Ali:2019lzf}        & (0.46$\pm$0.16)$\times 
10^{-6}$~\protect\cite{Strakovsky:2019bev} & (3.08$\pm$0.55)$\times 10^{-3}$~\protect\cite{Strakovsky:2019bev}\\
$\Upsilon(1S)$   & 521~\protect\cite{Guo:2021ibg}        & (0.37$\pm$0.04)$\times 
10^{-9}$ & (0.51$\pm$0.03)$\times 10^{-3}$ \\
\hline
\end{tabular}} \label{tbl:tab3}
\end{table*}
Figure~\ref{fig:fig3} illustrates the dramatic differences in the hadronic factors $h_{Vp} = \sqrt{b_1}$ (see Table~\ref{tbl:tab3}), as the slopes ($b_1$ from Eq.~(\ref{eq:eq2})) of the total cross sections at threshold as a function of $q$ vary significantly from $\omega$ to $J/\psi$ and now to $\Upsilon$.

Therefore, such a big difference in SLs of the vector-meson--proton systems is determined mainly by the hadronic factor $h_{Vp}$, and reflects a strong weakening of the interaction in the $\bar bb-p$ and $\bar cc-p$ systems compared to that of the light $\bar qq-p$ ($q = u, d$) configurations. The interaction in the $\bar ss-p$ has an intermediate strength that is manifested in an intermediate value of the $\phi p$ SL.

\begin{figure}[htb]
\begin{center}
\includegraphics[height=2.5in, keepaspectratio]{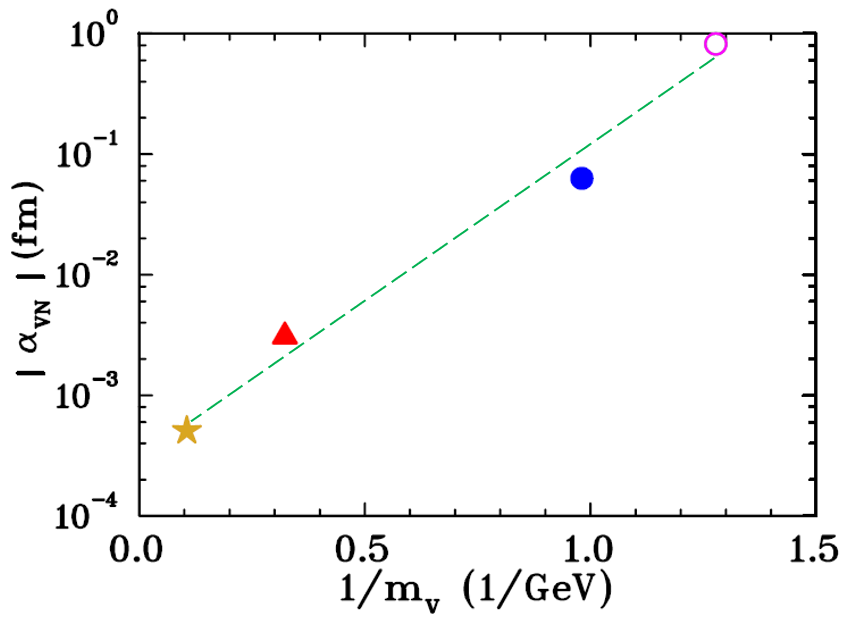}
\end{center}

\vspace{-7mm}
\caption{Comparison of the $|\alpha_{Vp}|$ SLs estimated from vector-meson photoproduction at     
        threshold vs. the inverse mass of the vector mesons.  The magenta open circle shows the analysis of the A2 $\omega$-meson data~\protect\cite{Strakovsky:2014wja}, the blue filled circle shows the analysis of the CLAS $\phi$-meson data~\protect\cite{Strakovsky:2020uqs}, the red filled triangle shows the analysis of the GlueX $J/\psi$-meson data~\protect\cite{Strakovsky:2019bev}, and the yellow filled star shows the analysis of the EIC and EicC $\Upsilon$-meson quasi-data~\protect\cite{Guo:2021ibg}. The green dashed line is hypotetical.} 
        \label{fig:fig4}
\end{figure}

The corresponding results for the scattering lengths are shown in Fig.~\ref{fig:fig4} as a function of the inverse vector-meson mass.  Starting from the $\phi $ meson and for higher masses, the scattering lengths are significantly smaller than the typical hadron size of 1~fm, indicating increasing transparency of the proton for these mesons. Moreover, our analysis shows almost linear exponential increase $|\alpha_{Vp}|\propto\exp(1/m_V)$ with increasing $1/m_V$. Actually, $p\to V$ coupling is proportional to the strong coupling $\alpha_s$ and the separation of the corresponding quarks. This separation (in zero approximation) is proportional to $1/m_V$, which may explain the above relation.

\section{Discussions and Outlook}
\label{sec:sum}

In summary, one can extend the relationship~(\ref{eq:eqa}) including the result for $\Upsilon$ (Table~\ref{tbl:tab3}):
\begin{equation}
        |\alpha_{\Upsilon p}| \ll |\alpha_{J/\psi p}| \ll |\alpha_{\phi p}| \ll |\alpha_{\omega p}|.
        \label{eq:eq8}
\end{equation}

The values of $|\alpha_{Vp}|$ for the heavy vector-mesons, as determined in the recent papers~\cite{Strakovsky:2014wja,Strakovsky:2020uqs,Strakovsky:2019bev} using the VMD model, are smaller than most of the theoretical predictions; see references in Refs.~\cite{Strakovsky:2014wja,Strakovsky:2020uqs,Strakovsky:2019bev}. As for the $\Upsilon$ case, we are not aware of any theoretical predictions for the $\Upsilon p$ scattering length except, related to that,  a non-relativistic potential result for the radius of the bottominium which is $r_\Upsilon = (0.14\pm 0.0014$)~fm~\cite{Satz:2006kba}. The same approach gives a radius of the charmonium of $r_{J/\psi} = (0.25\pm 0.0025$)~fm. Gryniuk and co-workers reported recently $|\alpha_{\Upsilon  p}| = (0.066\pm 0.001$)~fm and $|\alpha_{\Upsilon p}| = (0.016\pm 0.001$)~fm~\cite{Gryniuk:2020mlh} for two different subtraction constants. The latter results are based on the dispersive relations, taking some assumptions from their $J/\psi$ photoproduction model~\cite{Gryniuk2016}, and extrapolating the high-energy measurements at $W \approx 100$~GeV down to the threshold.

The smallness of the scattering lengths, as extracted with the help of the VMD model, can be related to the ``young age'' of the vector mesons participating in the interaction with the proton as introduced by Feinberg~\cite{Feinberg:1980yu}. For more quantitative estimate of the theoretical uncertainty related to the VMD model, we refer to the paper by Boreskov and Ioffe~\cite{Boreskov:1976dj}. They evaluated the cross section of $J/\psi$ photoproduction in a peripheral model and found a strong energy dependence because the non-diagonal process $\gamma p\to J/\psi p$ must have larger transfer momenta versus the elastic scattering $J/\psi p\to J/\psi p$. This result is in a violation of VMD by a factor of 5 or so. In the case of the $\Upsilon$ meson, we expect the disagreement to be even worse. Boreskov and co-workers showed that a fluctuation of a photon into open charm particles is preferable than into charmonium~\cite{Boreskov:1992ur}. Later on, it was shown by Kopeliovich and co-workers that the open-charm cross section is larger than the charmonium one by a factor of 10 or more~\cite{Hufner:2000jb,Floter:2007xv}. In addition, in Eq.~(\ref{eq:eq3}), we did not include a factor introduced in the VMD model in Ref.~\cite{Kubarovsky:2015aaa}, which takes into account the difference between polarization degrees of freedom in the $\gamma p\to J/\psi p$ and $pJ/\psi\to pJ/\psi$ reactions. Such a factor that equals 2/3 at threshold for the $S$-wave has not been used in the previous analysis of the scattering lengths; we consider it as a systematic uncertainty related to the VMD model~\cite{Strakovsky:2019bev}. 

In the recent work of Ref.~\cite{Xu:2021mju} the effect of the VMD assumption was studied in the formalism of the Dyson-Schwinger equations which one can consider as an alternative interpretation 
of the ``young age'' effect in another (more formal) language.  Their result shows much dramatic effect, up to a factor of 50 overestimation of the cross sections when using VMD for the two heaviest vector mesons. Nevertheless, this translates into a factor of 7 for the scattering lengths and the overall dependence in Fig.~\ref{fig:fig4} remains similar.

Present and future experiments JLab, EIC, and EicC that are aimed to measure charmonium and bottomonium production on the proton and nuclei will allow further studies of $J/\psi N$ and $\Upsilon N$ interactions and will allow access to a variety of other interesting physics aspects that are present in the near-threshold region. Further studies on both nucleons and nuclei in heavy vector-meson photo- and electro-production will significantly extend our knowledge of the gluonic structure of the nuclear matter.

\vspace{5mm}
\section{Acknowledgments}

We thank Misha Ryskin for useful remarks and continuous interest in the paper, Yulia Furletova for the detector design for EIC, and Yuxun Guo for predictions of $\Upsilon$ photoproduction cross sections. This work was supported in part by the U.~S. Department of Energy, Office of Science, Office of Nuclear Physics under Awards No. DE--SC0016583 and Contract No. DE--AC05--06OR23177. 


\end{document}